\documentclass[aps,prl,onecolumn,showpacs]{revtex4}

\usepackage{amsmath,epsfig,amsfonts,epsfig,graphicx,
}

\begin{document}

\title{Intrinsic Energy Localization through Discrete Gap Breathers \\
in One-Dimensional Diatomic Granular Crystals}
\author{G. Theocharis$^{1}$, N. Boechler$^{1}$, P. G. Kevrekidis$^{2}$, S. Job$^{3,1}$, Mason A. Porter$^{4}$, and C. Daraio$^{1}$}
\affiliation{ 
$^1$ Graduate Aerospace Laboratories (GALCIT) California Institute of Technology, Pasadena, CA 91125, USA  \\
$^2$ Department of Mathematics and Statistics, University of Massachusetts, Amherst, MA 01003-4515, USA \\
$^3$ Supmeca, 3 rue Fernand Hainaut, 93407 Saint-Ouen, France \\
$^4$ Oxford Centre for Industrial and Applied Mathematics, Mathematical Institute, University of Oxford, OX1 3LB, UK 
}


\begin{abstract}

We present a systematic study of the existence and stability of discrete breathers that are spatially localized in the bulk of a one-dimensional chain of compressed elastic beads that interact via Hertzian contact.  The chain is diatomic, consisting of a periodic arrangement of heavy and light spherical particles.  We examine two families of discrete gap breathers: (1) an unstable discrete gap breather that is centered on a heavy particle and characterized by a symmetric spatial energy profile and (2) a potentially stable discrete gap breather that is centered on a light particle and is characterized by an asymmetric spatial energy profile. We investigate their existence, structure, and stability throughout the band gap of the linear spectrum and classify them into four regimes: a regime near the lower optical band edge of the linear spectrum, a moderately discrete regime, a strongly discrete regime that lies deep within the band gap of the linearized version of the system, and a regime near the upper acoustic band edge. We contrast discrete breathers in anharmonic FPU-type diatomic chains with those in diatomic granular crystals, which have a tensionless interaction potential between adjacent particles, and highlight in that the asymmetric nature of the latter interaction potential may
lead to a form of hybrid bulk-surface localized solutions.  

\end{abstract}



\maketitle


\section{Introduction}\label{sec1}

The study of granular crystals draws on ideas from condensed matter physics, solid mechanics, and nonlinear dynamics.  A granular crystal consists of a tightly packed, uniaxially compressed array of solid particles that deform elastically when in contact with each other.  One-dimensional (1D) granular crystals have been of particular interest over the past two decades because of their experimental, computational, and (occasionally) theoretical tractability, and the ability to tune the dynamic response to encompass linear, weakly nonlinear, and strongly nonlinear behavior by changing the amount of static compression \cite{nesterenko1,sen08,nesterenko2,coste97,Job05}.  Such systems have been shown to be promising candidates for many engineering applications, including shock and energy absorbing layers \cite{dar06,hong05,fernando,doney06,Melo06}, actuating devices \cite{dev08}, acoustic lenses \cite{Spadoni} and sound scramblers \cite{dar05,dar05b}.

Intrinsic localized modes (ILMs), which are also known as discrete breathers (DBs), have been a central theme for numerous theoretical \cite{camp04,flawil,macrev,Flach2007,aubrev,sietak,pa90,macaub,thermal,kenig09} and experimental studies \cite{Swanson1999,Schwarz1999,orlando1,Binder2000,Sato2004,photon,Xiepeyrard,sievers,morsch1,Cuevas09} for more than two decades. Granular crystals provide an excellent setting to investigate such phenomena further.  Recent papers have begun to do this, considering related topics, e.g., metastable breathers in acoustic vacuum \cite{mohan}, localized oscillations on a defect that can occur upon the incidence of a traveling wave \cite{Job}, and an investigation of the existence and stability of localized breathing modes induced by the inclusion of ``defect" beads within a host monoatomic granular chain \cite{Theo2009}; for earlier work see e.g. the reviews of~\cite{sen08,nesterenko1}. 

Very recently, we reported the experimental observation of DBs in the weakly nonlinear dynamical regime of 1D diatomic granular crystals \cite{Boechler2009}. In \cite{Boechler2009} we describe the characteristics of the DB to be a few number of particles oscillate with a frequency in the forbidden band (i.e., the gap) of the linear spectrum, with an amplitude which decreases exponentially from the central particle. We took advantage of a modulational instability in the system to generate these breathing modes, and found good qualitative and even quantitative agreement between experimental and numerical results. It is the aim of the present paper to expand on these investigations with a more detailed numerical investigation of the existence, stability and dynamics of DBs in a diatomic, strongly compressed granular chain. In this paper, we examine two families of DBs lying within the gap of the linear spectrum (or discrete gap breathers-DGBs). By varying their frequency, DGBs can subsequently
be followed as a branch of solutions. The family that is centered around a central light mass and has an asymmetric energy profile can potentially be stable sufficiently close to the lower optical band edge before becoming weakly unstable when continued further into the gap.  Other solutions, such as the family 
that is centered around a central heavy mass and has a symmetric energy profile seem to always be unstable.  We examine both light and heavy mass centered families using direct numerical simulations.  

The study of DGB has importance both for increasing understanding of the nonlinear dynamics of strongly compressed granular elastic chains, and for the potential to enable the design of novel enginnering devices.  For instance, in the past there have been several attempts to design
mechanical systems to harvest or channel energy from ubiquitous random vibrations and noise of mechanical
systems \cite{Jeon2005,Cottone2009}. However, a drawback of such attempts has been that the energy of ambient vibrations is distributed over a wide spectrum of frequencies. Our recent experimental observation of intrinsic (and nonlinear) localized modes in chains of particles \cite{Boechler2009} opens a new possible mechanism for locally trapping vibrational energy in desired sites and harvesting such long-lived and intense excitations directly (e.g., by utilizing piezo-materials \cite{Sodano2004}).

The remainder of this manuscript is structured as follows: We first report the theoretical setup 
for our investigations and discuss the system's linear spectrum.  We then give an overview of the families of the DGB solutions that we obtain and present a systematic study of their behavior, categorized into four regimes relating to the frequency and degree of localization of these solutions. 
Finally, we summarize our findings and suggest some interesting directions for future studies.



\section{Theoretical Setup}\label{sec2}

\subsection{Equations of Motion and Energetics}

\begin{figure}[tbp]
\begin{center}
\includegraphics[width=9cm,height=2cm,angle=0,clip]{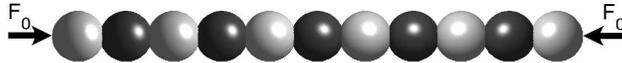}
\end{center}
\caption{Schematic of the diatomic granular chain. Light gray represents aluminum beads, and dark gray represents stainless steel beads.}
\label{setup}
\end{figure}

We consider a 1D chain of elastic solid particles, which are subject to a constant compression force $F_0$ that is applied to both free ends as shown in Fig. \ref{setup}. The Hamiltonian of the system is given by
\begin{equation} 
	H=\sum_{i=1}^{N}\left[\frac{1}{2}m_i\left(\frac{du_i}{dt}\right)^2+V(u_{i+1}-u_{i})\right]\,,
\label{Hamilton}
\end{equation} 
where $m_i$ is the mass of the $i$th particle, $u_i=u_i(t)$ is its displacement from the equilibrium position in the initially compressed chain, and $V(u_{i+1}-u_{i})$ is the interaction potential between particles $i$ and $i+1$.

We assume that stresses lie within the elastic threshold (in order to avoid plastic deformation of the particles) and that the particles have sufficiently small contact areas and velocities, so that we can make use of tensionless, Hertzian power-law interaction potentials.  To ensure that the classical ground state, for which $u_i={\dot u_{i}}=0$, is a minimum of the energy $H$, we also enforce that the interaction potential satisfies the conditions $V'(0) = 0$, $V''(0) > 0$.  The interaction potential can thus be written in the following form \cite{sen08,MacKay99}:
\begin{equation} 
V(\phi_{i})=\frac{1}{n_{i}+1}\alpha_{i,i+1}[\delta_{i,i+1}+\phi_{i}]_{+}^{n_{i}+1}-\alpha_{i,i+1}\delta_{i,i+1}^{n_{i}}\phi_{i}-\frac{1}{n_{i}+1}\alpha_{i,i+1}\delta_{i,i+1}^{n_{i}+1}\,,
\label{HertzPot}
\end{equation} 
where $\delta_{i,i+1}$ is the initial distance (which results from the static compression force $F_0$) between the centers of adjacent particles.  Additionally, $\phi_{i}=u_{i}-u_{i-1}$ denotes the relative displacement, and $\alpha_{i,i+1}$ and $n_{i}$ are coefficients that depend
on material properties and particle geometries.  The bracket $[s]_+$ of Eq.~(\ref{HertzPot}) takes the value $s$ if $s > 0$ and the value $0$ if $s \leq 0$ (which signifies that adjacent particles are not in contact).  

The energy $E$ of the system can be written as the sum of the energy densities $e_{i}$ of each of the particles in the chain:
\begin{eqnarray}
	E &=& \sum_{i=1}^{N}e_i, \nonumber \\
e_i &=& \frac{1}{2}m_i{\dot u_i}^2+\frac{1}{2}[V(u_{i+1}-u_i)+V(u_{i}-u_{i-1})]\,.
\label{energy}
\end{eqnarray}
In this paper, we focus on spherical particles. For this case, the Hertz law yields
\begin{equation} 
	\alpha_{i,i+1}=\frac{4E_iE_{i+1}\sqrt{\frac{R_{i}R_{i+1}}{R_i+R_{i+1}}}}{3E_{i+1}(1-\nu_i^2)+3E_{i}(1-\nu_{i+1}^2)}\,, \qquad n_{i}=\frac{3}{2}\,,
\label{coeff}
\end{equation} 
where the $i$th bead has elastic modulus $E_i$, Poisson ratio $\nu_i$, and radius $R_i$.  Hence, a 1D diatomic chain of $N$ alternating spherical particles can be modeled by the following system of coupled nonlinear ordinary differential equations:
\begin{equation}
	m_{i}\ddot{u}_i = A[\delta_{0}+u_{i-1} - u_{i}]_{+}^{3/2} - A[\delta_{0}+u_{i} - u_{i+1}]_{+}^{3/2}\,, 
\label{model}
\end{equation}
where  $A=\alpha_{i,i+1}=\frac{4E_1E_{2}\left(\frac{R_1R_{2}}{R_1 + R_{2}}\right)^{1/2}}{3\left(E_{2}(1-\nu_1^2)+(E_{1}(1-\nu_{2}^2)\right)}$,  $\delta_{0}=\delta_{i,i+1}=\left(\frac{F_0}{A}\right)^{2/3}$, and we recall that $F_0$ is  the static compression force. The particle masses are $m_{2i-1}=m$ and $m_{2i}=M$ for $i \in \{1,\cdots,N\}$.  By convention, we will take $M$ to be the larger of the two masses and $m$ to be the smaller of the two masses. 
The equations of motion for the beads at the free ends are
\begin{align}
	m_1\ddot{u}_1&=F_0-A[\delta_{0}-(u_2-u_1)]_+^{3/2}\,,\label{e:leftBC}\\
	m_N\ddot{u}_N&=A[\delta_{0}-(u_N-u_{N-1})]_+^{3/2}-F_0\,. \label{e:rightBC}
\end{align}

\subsection{Weakly Nonlinear Diatomic Chain}

If the dynamical displacements have small amplitudes relative to those due to the static compression $(|\phi_i|<\delta_{0})$, we can consider the weakly nonlinear dynamics of the granular crystal. It is the interplay of this weak nonlinearity with the discreteness of the system that allows the existence of the DGB. To describe this regime, we take a power series expansion of the forces (up to quartic displacement terms) to yield the, so-called, $K_2-K_3-K_4$ model:
\begin{equation}
	m_{i}\ddot{u}_i=\sum_{j=2}^{4}K_{j}\left[(u_{i+1}-u_{i})^{j-1}-(u_{i}-u_{i-1})^{j-1} \right]\,,
\label{K2K3K4}
\end{equation}
where $K_2=\frac{3}{2}A^{2/3}F_0^{1/3}$ is the linear stiffness, $K_3=-\frac{3}{8}A^{4/3}F_{0}^{-1/3}$, and $K_4=\frac{3}{48}A^{2}F_{0}^{-1}$.  


\subsection{Linear Diatomic Chain} 

For dynamical displacements with amplitude much less than the static overlap $(|\phi_i|\ll\delta_{0})$, we can neglect the nonlinear $K_3$ and $K_4$ terms from
Eq.~(\ref{K2K3K4}) and compute the linear dispersion relation of the system \cite{Herbold09}. The resulting diatomic chain of masses coupled by harmonic springs
is a textbook model for vibrational normal modes in crystals \cite{kittel}. Its
dispersion relation contains two branches (called \emph{acoustic} and \emph{optical}). At the edge of the first Brillouin zone---i.e., at wave number $k= \frac{\pi}{2\Delta_0}$, where $\Delta_0=R_a+R_b-\delta_0$ is the equilibrium distance between two adjacent beads---the linear spectrum possesses a gap between the upper cutoff frequency $\omega_{1}=\sqrt{2K_2/M}$ of the acoustic branch and the lower cutoff frequency $\omega_{2}=\sqrt{2K_2/m}$ of the optical branch. 
The upper cutoff frequency of the optical band is located at $\omega_{3}=\sqrt{2K_2(1/m+1/M)}$.

In addition to acoustic and optical modes, the diatomic semi-infinite harmonic chain also supports a gap mode, provided the existence of a light particle at the surface and the use of free boundary conditions. This mode is localized at the surface (i.e., at the first particle) and its displacements have the following form \cite{Wallis57}:
\begin{align}
	u_{2k+1}  &= B(-1)^{k}\left(\frac{m}{M}\right)^{k}e^{j\omega_s t}\,, \label{e:left1BC}\\
	u_{2k+2} &= B(-1)^{k+1}\left(\frac{m}{M}\right)^{k+1}e^{j\omega_s t}\,, \label{e:right1BC}
\end{align}
with $k\geq 0$, frequency $\omega_s=\sqrt{K_2(1/m+1/M)}$ in the gap of the linear spectrum and B an arbitrary constant. Thus, the surface mode decays exponentially with a characteristic decay length of
\begin{equation}\label{charlength}
	\xi = 2\Delta_0/\ln(M/m)\,. 
\end{equation}

A standard derivation of the surface mode is given in Ref.~\cite{Cottam89}, while a simple physical explanation of its existence and characteristics can be found in Ref.~\cite{Allen99}.  The latter is summarized as follows: Adjacent pairs vibrate in such a way that the connecting spring is not stretched. Thus each pair experiences no force from any other particle and is decoupled from the rest of the chain. The resulting decoupled pairs oscillate with $\omega_s$.

This particular mode with frequency in the band gap, localized around the surface, proves to have a nonlinear counterpart and be very closely related to the DGB in the strongly discrete regime as we describe in later sections. 


\subsection{Experimental Determination of Parameters}

In our experiments from \cite{Boechler2009} and numerical simulations, we consider a 1D diatomic granular crystal with alternating aluminum spheres ($6061$-T$6$ type, radius $R_a=9.53$~mm, mass $m = m_a=9.75$~g, elastic modulus $E_a=73.5$~GPa, Poisson ratio $\nu_a=0.33$) and stainless steel spheres ($316$ type, $R_b=R_a$, $M = m_b=28.84$~g, $E_b=193$~GPa, $\nu_b=0.3$). The values of $E_{a,b}$ and $\nu_{a,b}$ that we report are standard specifications~\cite{ElasticProperties}.  In Ref.~\cite{Boechler2009}, we experimentally characterized the linear spectrum of this diatomic crystal and we calculated the particle's effective parameter $A=7.04$ N/$\mu$m$^{3/2}$.  Using this value with the theoretical formulas above, we calculate the cutoff frequencies and the surface mode frequency.  We summarize these results for a static load of $F_0=20$~N in Table~\ref{table_th_vs_exp_linear}.  
For the rest of the manuscript we use this experimentally determined effective parameter A in our numerical analyses. 

\begin{table}[h]
\begin{tabular}{|c|c|c|c|c|}
\hline          A [N/$\mu$m$^{3/2}$] &$f_1$ [kHz] & $f_2$ [kHz] & $f_3$ [kHz] & $f_s$ [kHz]    \\ 
\hline          7.04                 &5.125         & 8.815         & 10.20   & 7.21            \\ 
\hline 
\end{tabular}
\caption{\label{table_th_vs_exp_linear} Calculated cutoff frequencies (based on the experimentally-obtained coefficient $A$ \cite{Boechler2009}) under a static compression of $F_0=20$~N.}
\end{table}

\begin{figure}[tbp]
\begin{center}
\includegraphics[width=8.5cm,height=6cm,angle=0,clip]{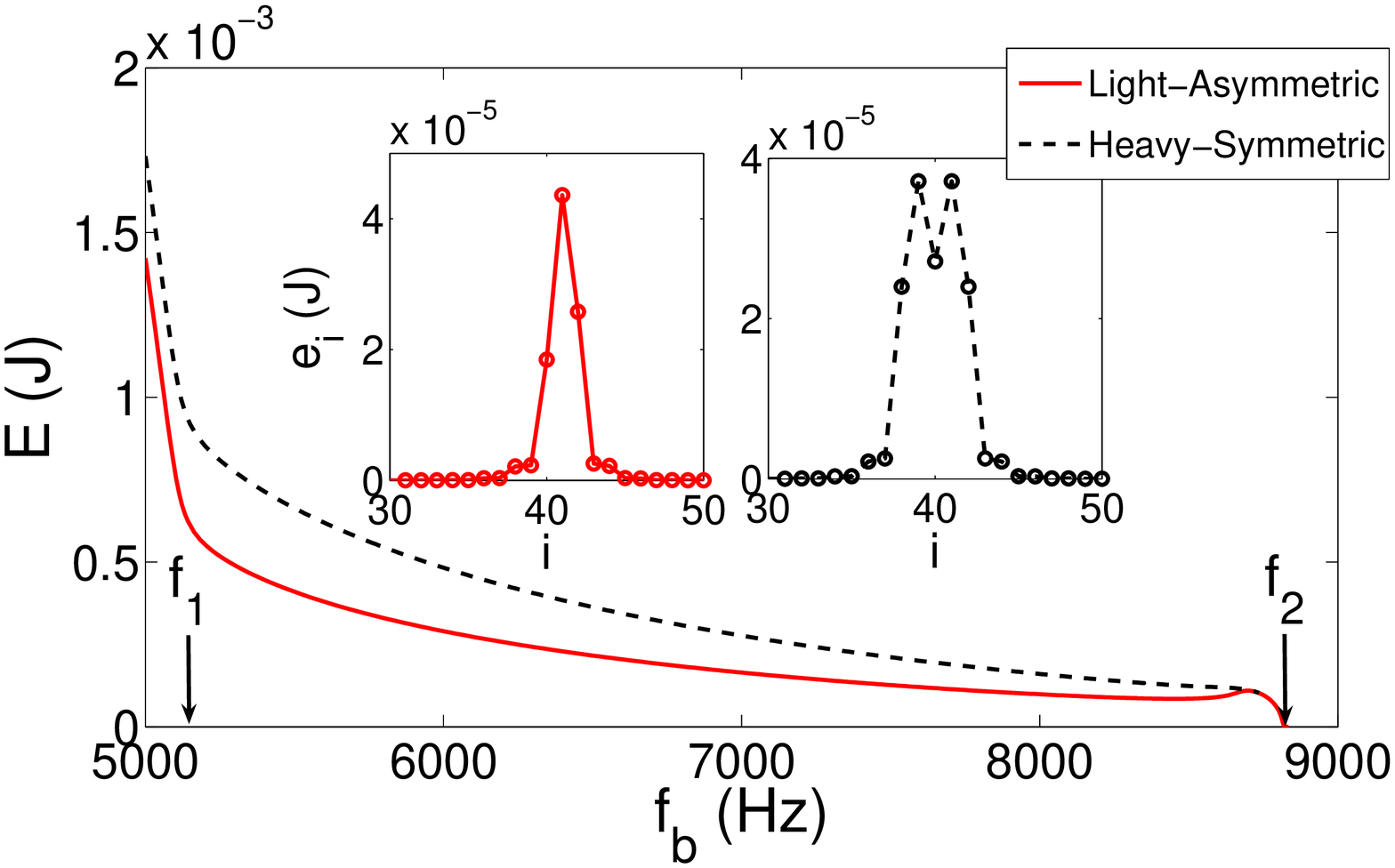}
\end{center}
\caption{(Color online) Energy of the two families of discrete gap breathers (DGBs) as a function of their frequency $f_b$. The inset shows a typical example of the energy density profile of each of the two modes at $f_b = 8000$ Hz. 
}
\label{En_fb}
\end{figure}


\section{Overview of Discrete Gap Breathers}\label{sec3}
 

\subsection{Methodology}

As the equations of motion (\ref{model}) are similar to the equations of motion in the FPU-type problem of \cite{focus,ford}, we accordingly recall relevant results. 
A rigorous proof of the existence of DBs in a diatomic FPU chain with alternating heavy and light masses (which is valid close to the $m/M \rightarrow 0$ limit) can be found in Ref.~\cite{Livi1997}. Information about the existence and stability of DBs in the gap between the acoustic and optical band of an anharmonic diatomic lattice can be found in Refs.~\cite{Kiselev1993,Maniadis2003,FPUDiatomic,Huang1998}.  At least two types of DGBs are known to exist, and (as we discuss below) both can arise in granular chains.

We conduct numerical simulations of a granular chain that consists of $N = 81$ beads (except where otherwise stated) and free boundaries.  In order to obtain DGB solutions with high precision, we solve the equations of motion (\ref{model}, \ref{e:leftBC}, \ref{e:rightBC}) using Newton's method in phase space.  This method is convenient for obtaining DGB solutions with high precision and for studying their linear stability.  Additionally, we can obtain complete families of solutions using parameter continuation; one chooses system parameters corresponding to a known solution and subsequently changes the parameters using small steps.  For a detailed presentation of the numerical methods, see Ref.~\cite{Flach2007} and references therein. 


\subsection{Families of Discrete Gap Breathers}

The initial guess that we used to identify the DGB modes is the lower optical cutoff mode, obtained by
studying the eigenvalue problem of the linearization of Eq.~(\ref{model}). Such a stationary profile
(with vanishing momentum) is seeded in the nonlinear Newton solver to obtain the relevant breather-type
periodic orbits. Continuation of this lower optical mode inside the gap
allows us to follow one family of DGB solutions. By examining the energy density profiles of these solutions,
we observe that they are characterized by an asymmetric localized distribution of the energy centered at the
central light bead of the chain (see the left inset of Fig.~\ref{En_fb}). We will henceforth refer to this family of
solutions with the descriptor \emph{LA} (light centered-asymmetric energy distribution).

For frequencies deep within  the band gap, the DGB is not significantly affected by the boundary conditions since it extends only over
few particles. Thus, its pinning site may be placed at any light bead of the chain, not only the central one. However, as the
frequency of the DGB solution approaches the lower optical cutoff, the solution becomes more and more extended and
the boundaries come into play.
For instance, using an initial guess of a LA-DGB solution deep within the gap, shifted by a unit cell to the left, we performed a
continuation throughout the frequency gap, and obtained a similar family of LA-DGB (but shifted by one unit cell). It is interesting to 
note that this family of breathers does not bifurcate from the optical band, as the family of the LA-DGB centered at the
central light bead does, but rather ceases to exist at $f_b\approx 8755$ Hz.

These families of DGB solutions, centered at light beads, are not the only ones that our system supports. We were able
to trace a second type of family as well. The energy density profiles for solutions in this second family are symmetric and
centered on a heavy bead (see the right inset of Fig.~\ref{En_fb}). We will call this family of solutions \emph{HS} (heavy
centered-symmetric energy distribution). The seed for this solution (as will be described in further detail) may be obtained by
perturbing the LA-DGB along an eigenvector associated with translational symmetry. A continuation of this family of solutions
can be performed as well.
Increasing the frequencies towards the optical cutoff band, we found that the HS-DGB family of solution, centered at the central
heavy bead, also ceases to exist at $f_b\approx 8755$ Hz. This branch of DGBs is linearly unstable and at that frequency
experiences a saddle-center bifurcation with the LA-DGB branch of solutions shifted one unit cell from the middle of the chain,
and thus both families of DGBs dissapear. The bifurcation point depends on the length of the system and specifically, the larger the
system size, the closer the frequency of the bifurcation to the lower optical cutoff frequency.

The above phenomenology can be generalized for all the shifted families of DGB solutions (namely LA and HS-DGB with different pinning sites).
Approaching the optical band, consecutive pairs of HS-DGB and LA-DGB solutions collide and disappear.  This cascade of pairwise saddle-center
bifurcations occurs  closer to the optical band edge, the further away from the chain boundary the pair of LA and HS-DGBs is centered.
Only one branch of solution, the LA-DGB solution centered at the central light bead, survives and ends at the linear limit of the optical lower cutoff edge.
In this paper, we will focus on two families of DGB solutions. The LA-DGB centered at the central light bead and the HS-DGB
centered at the central heavy bead. 

In Fig.~\ref{En_fb}, we show the dependence of the breather's energy on its frequency
and (in the insets) examples of the spatial energy profile of these two different families of DGB solutions (with frequency $f_b=8000$ Hz)
that the system supports. As one can observe in the energy diagram, the energies of the two solutions are very close around $f_b\approx 8755$ Hz,
while the energy of the LA-DGB approaches zero  as the frequency of the breather approaches the optical lower cutoff frequency $f_2$. 
In contrast, when the frequency of the breather approaches the acoustic upper cutoff frequency $f_1$, the energies of the solutions grow rapidly. 
As we discuss below, this arises from the resonance of the DGB  with the linear acoustic upper cutoff mode.

Finally, as briefly discussed in Ref.~\cite{Boechler2009}, the energy of the LA-DGBs appears to have turning points (i.e., points at which $dE/df_b=0$) at $f \approx 8480 Hz$ and $f \approx 8700 Hz$.  These turning points are directly associated with the real instability that the branch of LA-DGB solutions has in that frequency regime. This has also been observed in  binary discrete nonlinear Schr{\"o}dinger (DNLS) models with 
alternating on-site potential \cite{Gorbach2004} and diatomic Klein-Gordon chains \cite{Gorbach2003}.

\begin{figure}[tbp]
\begin{center}
\includegraphics[width=8.5cm,height=6cm,angle=0,clip]{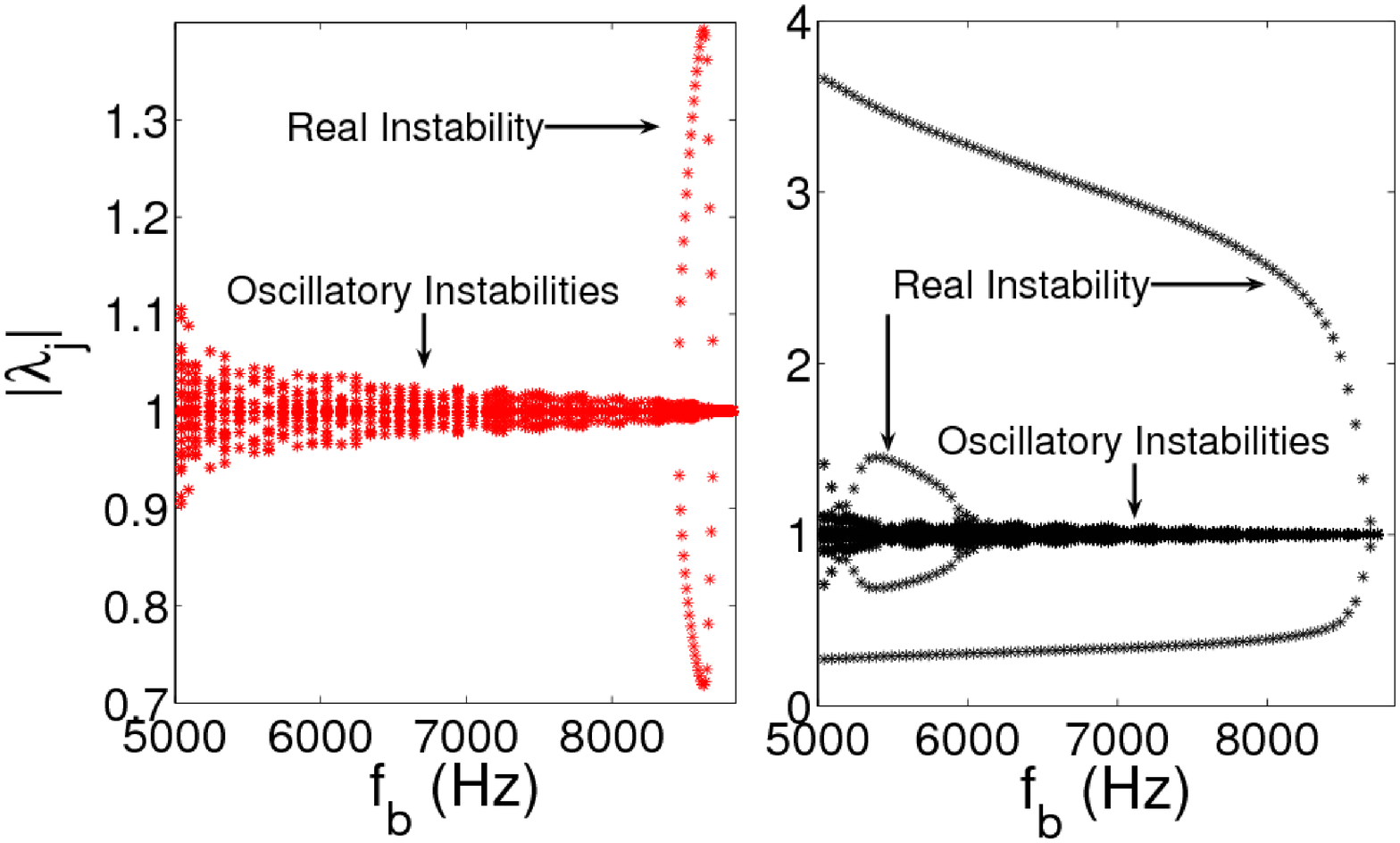}
\end{center}
\caption{(Color online) Magnitude of the Floquet mulitpliers as a function of DGB frequency $f_b$ for the DGB with a light centered-asymmetric energy distribution (LA-DGB; left panel) and for the DGB with a heavy-centered symmetric energy distribution (HS-DGB; right panel). 
}
\label{St_fb}
\end{figure}

 
\subsection{Stability Overview} 

In order to examine the linear stability of the obtained solutions, we compute their Floquet multipliers $\lambda_{j}$ \cite{Flach2007,aubrev}. If all of the multipliers $\lambda_{j}$ have unit magnitude, then the DGB is linearly stable for our Hamiltonian dynamical model.  Otherwise, it is subject to either real or oscillatory instabilities, for which the modulus of the corresponding unstable eigenvector grows exponentially as a function of time. It is important to note that two pairs of Floquet multipliers are always located at $(1,0)$ in the complex plane. One pair, corresponding to the phase mode, describes a rotation of the breather's aggregate phase.  The second pair arises from the conservation of the total mechanical momentum, an additional integral of motion that arises in FPU-like chains with free ends \cite{Flach2007}.

In Fig.~\ref{St_fb}, we show the resulting stability diagram for both families of DGB solutions. Strictly speaking, the DGBs are linearly stable only for $f_b$ very close to $f_2^{\text{exp}}$. For all other frequencies, both families of the DGBs exhibit either real or oscillatory instabilities \cite{aubrev}.  Real instabilities are connected to the collision of a pair of Floquet multipliers---the eigenvectors of which are spatially localized---at the points $(+1,0)$ or $(-1,0)$ on the unit circle. 
These instabilities are associated with growth rates that are typically independent of the size of the system (i.e., of the number of particles in the chain).  On the other hand, oscillatory instabilities can arise due to the collision of either two Floquet multipliers associated with spatially extended eigenvectors or one multiplier associated with a spatially extended eigenvector and another associated with a spatially localized one. Such collisions require that Krein signatures of the associated colliding eigevectors are opposite \cite{aubrev}. From a physical perspective, the Krein signature is the sign of the Hamiltonian energy that is carried by the corresponding eigenvector \cite{Skryabin2001}. Oscillatory instabilities can occur at any point on the unit circle.
 
The first type of oscillatory instability, which arises from the collision of two spatially extended eigenvectors, is known to be a finite-size effect.  As discussed in Ref.~\cite{Finitesize}, the strength of such instabilities should depend on the system size.  In particular, when the size of the system is increased, the magnitude of such instabilities weakens uniformly.  Simultaneously, the number of such instabilities increases with system size due to the increasing density of colliding eigenvalues. Eventually, these instabilities vanish in the limit of an infinitely large system.

The second type of oscillatory instability, which arises from a collision of a spatially localized eigenvector with a spatially extended eigenvector, occurs when an internal mode of the DGB (i.e., a localized eigenvector) enters the band of extended states associated with the phonon spectrum of the system(such extended eigenvectors are only sligthly modified due to the presence of the DGB). This kind of oscillatory instability does not vanish in the limit of an infinitely large system and is directly connected with Fano-like resonant wave scattering by DGBs (see e.g. Ref.~\cite{Flach2005} for the monoatomic FPU case).


\section{Four Regimes of Discrete Gap Breathers: Existence and Stability}\label{sec4}

 
\subsection{Overview of Four Dynamical Regimes}

The purpose of this section is to qualitatively categorize the two families of DGB into four regimes ({\it (I) close to the optical band, (II) moderately discrete, (III) strongly discrete,} and {\it (IV) close to and slightly inside the acoustic band}) according to DGB characteristics such as the maximum relative displacement and the localization length of the DGB, denoted by $l$ (DGBs are localized vibrational modes with amplitude which decays exponentially as $\exp(-|i|/l)$). Recalling that $\xi$ is the localization length of the linear surface mode, we find that this length, $\xi$, consists of a lower bound for the localization length $l$ of both families of DGBs.

\begin{figure}[tbp]
\begin{center}
\includegraphics[width=12cm,height=6cm,angle=0,clip]{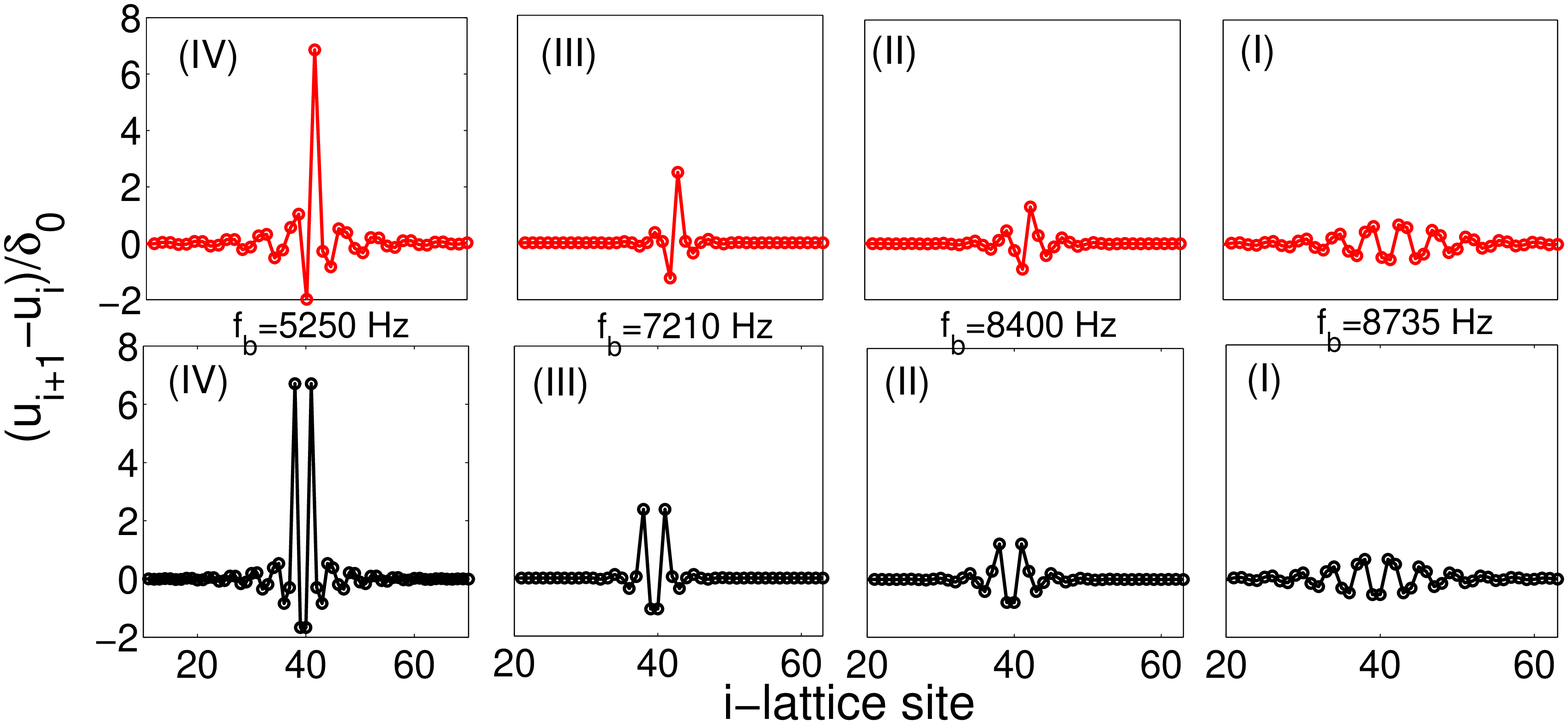}
\end{center}
\caption{(Color online) 
Top panels: Four  typical examples of the relative displacement profile of LA-DGB solutions, each one from a different dynamical regime.
Bottom panels: As with the top panels, but for HS-DGB solutions.}
\label{MaxF_fb} 
\end{figure}

In the top panels of Fig.~\ref{MaxF_fb}, we show typical examples of the relative displacement profiles of LA-DGB solutions (each of which occurs in a different regime of the band gap). We similarly show four typical HS-DGB solutions (at the same frequencies) in the bottom panels. In Table \ref{regimes}, we summarize the characteristics of the DGB solutions in the four regimes.

\begin{table}[h]
\begin{tabular}{|c|c|c|c|c|}
\hline & Regime (I) & Regime (II) & Regime (III) & Regime (IV) \\
\hline                                               &   Near Optical & Moderately Discrete & Strongly Discrete &  Near Acoustic   \\ 
\hline  $\max\frac{|u_{i}-u_{i+1}|}{\delta_{0}}$      &   $<1$             & $\gtrsim 1$         & $\gg1$          & $\gg1$             \\ 
\hline  localization length                          &   $l\gg \xi$       & $l>\xi$             & $l\gtrsim \xi$  & $l\gg \xi$         \\
\hline 
\end{tabular}
\caption{\label{regimes} Characteristics of the DGBs in the four different regimes. }
\end{table}

In addition to the differences in amplitude and localization length, each regime displays some uniquely interesting characteristics. The {\it close to the optical} regime contains the only strictly linearly stable modes (LA-DGB). The {\it moderately discrete} regime includes (but is larger than) the aforementioned region of strong instability for the LA-DGB, and is the region within which our experimentally observed LA-DGB \cite{Boechler2009} falls. The {\it strongly discrete} regime shows a change in spatial displacement profile (in both families) with respect to the other regimes, which (as will be discussed) is connected to large time loss of contact between adjacent beads (gap openings) and the existence of gap surface modes, and is unique to our tensionless contact potential. The {\it close to and slightly inside the acoustic band} regime shows resonances with the upper acoustic band edge, and (for the HS-DGB) a resulting period doubling bifurcation. 

We now continue, by conducting a detailed investigation of both DGB families in the four different regimes.


\subsection{Region (I): Close to the optical band ($f_b\lesssim f_2$)}

Regime (I) is located very close to the lower optical band edge of the linear spectrum. 
As we mention above, the HS-DGB family of solutions starts to exist from $f_b\approx 8755$ Hz and below, while the LA-DGB family of solutions
is initialized from the linear limit at the lower edge of the optical band.
In this regime,
both DGB solutions are characterized by a localization length $l$ that is much larger than the  characteristic localization length $\xi$ of the surface mode.  The DGB spatial profiles have the form of the spatial profile of the optical cutoff mode and the LA-DGB family is linearly stable. Moreover, as indicated by the relative displacements of the solutions in regime (I), we can also conclude that $\max\frac{|u_{i}-u_{i+1}|}{\delta_{0}}<1$, so the dynamics 
of the system is weakly nonlinear and the adjacent particles are always in contact (gaps do not open between particles). It should be noted that despite the linear stability, this similarity in frequency and spatial profile to the lower optical cutoff mode could make this regime of DGB difficult to observe experimentally and differentiate from the linear mode.   

In this regime, since both DGB solutions are characterized by a small amplitude and a large localization length $l$,
the effect of the discreteness is expected to be weak. Continuous approximation techniques (see, for example, Ref.~\cite{Huang1998}) have revealed that the dynamics of the envelope of the solutions close to the optical band is described by a focusing nonlinear Schr{\"o}dinger (NLS) equation.  Hence, the two types of DGBs can be viewed as discrete analogs of the asymmetric gap solitons that are supported by the NLS equation that is obtained in the asymptotic limit.

In this regime, due to the weak effect of the discreteness for $f_b\lesssim f_2$, a third pair of
Floquet multipliers appears in the vicinity of the point $(+1,0)$ on
the unit circle.  This is in addition to the two previously discussed pairs of Floquet multipliers relating to the phase mode and conservation of momentum. This third mode is the mode associated  with the {\it breaking} of the continuous translational symmetry (i.e., there is a discrete translational symmetry/invariance in the limit of small lattice spacing). The associated Floquet multiplier is of particular interest, as it has been associated with a localized mode called a ``translational" or ``pinning" mode \cite{aubrev}. Perturbing the LA-DGB solution along this corresponding Floquet eigenvector enables us to obtain the HS-DGB family of solutions (essentially translating light mass centered DGB by one site to a heavy centered DGB).


\subsection{Region (II): Moderately Discrete Regime}

We call regime (II) {\it moderately discrete}. In this regime, the localization length $l$ of the two DGB solutions is smaller than 
in the case of regime (I), so the effect of discreteness becomes stronger. In regime (II), we find that $\frac{|u_{i}-u_{i+1}|}{\delta_{0}}\gtrsim 1$ near the central bead. As a result this regime also has a larger kink-shaped distortion of the chain (i.e., displacement differential between
the left and right ends of the chain). This kink shaped distortion, which is visible in panel (a) of Fig. 
\ref{HA_8600}, is a static mutual displacement of the parts of the chain separated by the DGB, and
is a characteristic of DB solutions in anharmonic lattices that are described by asymmetric interparticle potentials (see for exmaple \cite{Kiselev1993,Huang1998}). The larger amplitude also translates into a gap opening
for the contact(s) of the central particle for a small amount of time. The response of the system can be considered as strongly nonlinear near the central bead of the breather and weakly nonlinear elsewhere. 

As the frequency is decreased throughout this regime, the effect of the discreteness
becomes stronger and the previously discussed ``pinning" mode moves away from the point $(+1,0)$ on the unit
circle; it moves along the unit circle for the LA-DGB solution and along the real axis for the HS-DGB solution (which causes a strong real instability in the latter case). For $8450 \mbox{ Hz } < f_b < 8700 \mbox{ Hz}$ both DGB families are subject to strong harmonic (and real) instability. However for $f_b< 8400 \mbox{ Hz}$, the LA-DGB is subject only to weak oscillatory instabilities whereas the HS-DBG maintains the real instability. It is in this regime, below the strong instability frequency region that we categorize the type of LA-DGB found experimentally in \cite{Boechler2009}. 

We discuss both HS and LA-DGB modes of this regime in further detail in the following sections.


\subsubsection{HS Discrete Gap Breather (HS-DGB)}

\begin{figure}[htbp]
\begin{center}
\includegraphics[width=8.5cm,height=7.5cm,angle=0,clip]{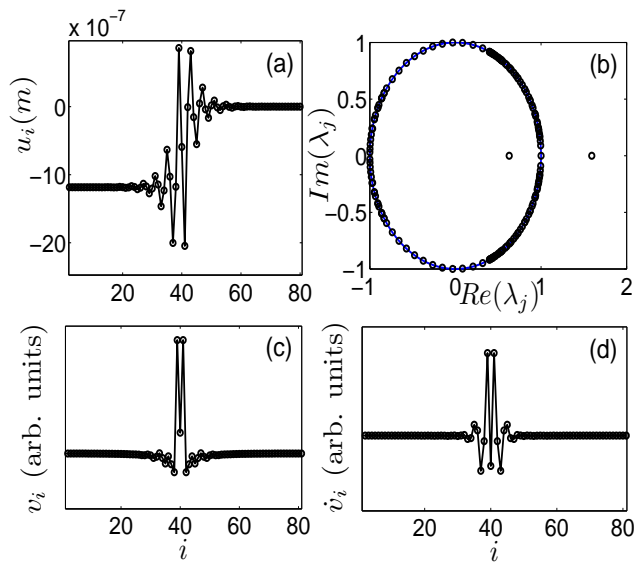}
\end{center}
\caption{(a) Spatial profile of an HS-DGB with frequency $f_b=8600$ Hz. (b) Corresponding locations of Floquet multipliers $\lambda_{j}$ in the complex plane. 
We show the unit circle to guide the eye. Displacement (c) and velocity components (d) of the Floquet
eigenvectors associated with the real instability.}
\label{HA_8600}
\end{figure}

In Fig.~\ref{HA_8600}(a), we show the spatial profile of an example HS-DGB solution with frequency $f_b=8600$ Hz located in the {\it moderately discrete} regime.  Observe in the stability diagram in panel (b) that one pair of Floquet multipliers has abandoned the unit circle and is 
positioned along the real axis. This strong instability (with a real multiplier) is caused by a localized Floquet eigenvector (the pinning mode).  We plot the displacement and velocity components of this eigenvector in panels (c) and (d), respectively.  This localized pinning mode is symmetric and centered at a heavy particle. Perturbing the HS-DGB along this unstable eigenvector deforms it in the direction of the LA-DGB.

\begin{figure}[htbp]
\begin{center}
\includegraphics[width=8.5cm,height=6cm,angle=0,clip]{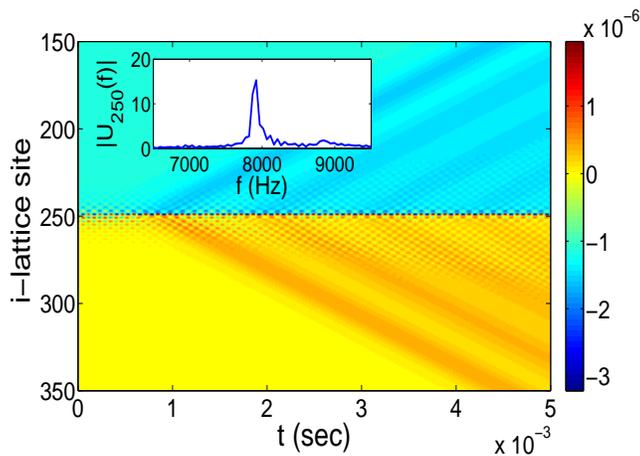}
\end{center}
\caption{(Color online) Spatiotemporal evolution (and transformation into $f_b \approx 7900$ Hz LA-DGB) of the displacements of
a HS-DGB summed with the pinning mode and initial $f_b=8600$ Hz.  Inset: Fourier transform of the center particle. 
}
\label{Inst_HA_8600}
\end{figure}

To reveal the effect of this instability (pinning mode) and elucidate the transition between HS and LA-DGB, we perform numerical integration of the original nonlinear equations of motion (\ref{model}) using as an initial condition the sum of the unstable HS-DGB mode and the pinning mode.  In order to reduce the reflecting radiation from the boundaries, we use a (rather large) chain that consists of $N = 501$ particles. The HS-DGB performs a few localized oscillations up to 
times of about $5T$ (where $T$ is the period of the solution);
then, it starts 
to emit phonon waves and eventually is transformed into a LA-DGB.  By performing a Fourier transform of the displacements of the center particle (see the inset of 
Fig.~\ref{Inst_HA_8600}), we find that the frequency of the transformed 
LA-DGB is $f_b \approx 7900$Hz.

%


\subsubsection{LA Discrete Gap Breather (LA-DGB)}

We now discuss the LA-DGB branch of solutions in the {\it moderately discrete} regime.  Carefully monitoring the motion of the Floquet multipliers on the unit circle during parameter continuation, we observe that at $f \approx 8717$ Hz, a pair of Floquet multipliers leaves the phonon band that consists of the eigenstates that
are spatially extended. The corresponding eigenmode becomes progressively more localized as the frequency decreases \cite{Baesens}.  At $f \approx 8700$ Hz, it arrives at the point $(+1,0)$ on the unit circle, where it collides with its complex conjugate to yield a real instability (see the left panel of 
Fig.~\ref{St_fb}). This instability persists down to $f \approx 8450$ Hz. As indicated above, this real instability is directly associated with the turning points that arise from the frequency dependence of the energy.  As discussed also in Ref.~\cite{Gorbach2003}, this is reminiscent of the Vakhitov-Kolokolov instability of soliton solutions for the NLS equation \cite{Vakhitov73}.

\begin{figure}[htbp]
\begin{center}
\includegraphics[width=8.5cm,height=7.5cm,angle=0,clip]{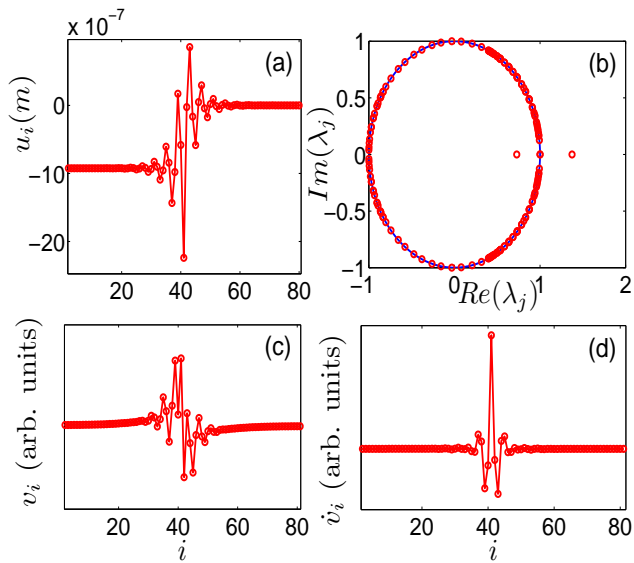}
\end{center}
\caption{(Color online)(a) Spatial profile of an LA-DGB with frequency $f_b=8600$Hz. (b) Corresponding locations of Floquet multipliers $\lambda_{j}$ in the complex plane. We show the unit circle to guide the eye. Displacement (c) and velocity (d) components of the Floquet
eigenvector associated with the real instability.}
\label{LS_8600}
\end{figure}

In Fig.~\ref{LS_8600}(a), we show the spatial profile of an example LA-DGB solution with frequency $f_b=8600$Hz (in the real instability region).  Observe in the stability diagram in panel (b) that one pair of Floquet multipliers has abandoned the unit circle and is
located along the real axis. This strong instability (arising from the real multiplier) is caused by a localized Floquet eigenvector.  We plot its displacement and velocity components in panel (c) and (d), respectively.  This mode is asymmetric and centered at a light particle.

\begin{figure}[htbp]
\begin{center}
\includegraphics[width=8.5cm,height=6cm,angle=0,clip]{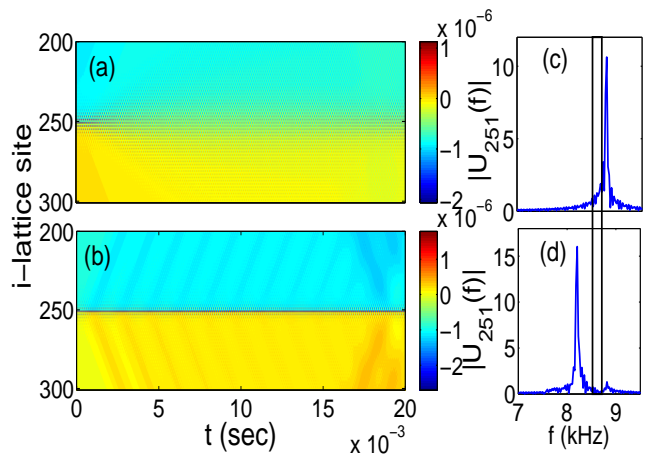}
\end{center}
\caption{(Color online) Spatiotemporal evolution of the displacements of a LA-DGB with $f_b=8600$ Hz when one (a) adds and (b) subtracts the unstable localized mode depicted in Fig.~\ref{LS_8600}(c).  Panel (c) shows the Fourier transform of the center particle for case (a), and panel (d) shows the same for case (b).
In panels (c,d), the two vertical lines enclose the regime of the frequencies in which the LA-DGB
exhibits the strong real instability.
}
\label{Inst_LS_8600}
\end{figure}

To reveal the effect of this instability, we perform numerical integration of the nonlinear equations of motion (\ref{model}).  As before, we use a large chain consisting of $N = 501$ particles in order to reduce the reflecting radiation from the boundaries. As one can observe in the top panels of Fig.~\ref{Inst_LS_8600}, the LA-DGB with frequency $f_b=8600$ Hz, which is subject to the strong real instability, is transformed into a linearly stable, more extended, LA-DGB with $f_b \approx 8800$ Hz when we add it to the solution the unstable Floquet eigenvector. On the other hand, as depicted in the bottom panels of Fig.~\ref{Inst_LS_8600}, we obtain an LA-DGB with $f_b \approx 8200$ Hz when we subtract the unstable eigenvector from the initial LA-DGB solution. Hence, it becomes apparent that depending on the
nature of the perturbation, the unstable LA-DGB can be ``steered''
towards higher or lower (more stable) oscillation frequencies
within the gap.

This is also important with respect to the experimental observation in \cite{Boechler2009}. As we've previously discussed, the DGB in the {\it nearly continuum} regime are very similar to the linear lower optical cutoff mode, and thus potentially difficult to detect. Following that regime, to this {\it moderately discrete} regime, both LA and HS-DGB are characterized by strong instabilities down to $f_b \approx 8400$ Hz. The HS-DGB continue to be characterized by a strong instability. As we showed in Fig.~\ref{Inst_HA_8600} the HS-DGB will transform to a LA-DGB below this region of strong instability. Furthermore, as we showed in Fig.~\ref{Inst_LS_8600} a LA-DGB in the strong instability region can transform to a LA-DGB of frequency either above or below the region of strong instability. It is natural then, as a more stable solution that is of significantly distinct profile, that the DGB observed in \cite{Boechler2009} is a LA-DGB in the {\it moderately discrete} regime with $f_b \approx 8280$ Hz. Although other effects such as dissipation will also play a role into the observability of the DGB modes, the stability and structural analysis can be the beginnings of a guide for observability in practice.  


\subsection{Region (III): Strongly Discrete Regime ($f_1\ll f_b \ll f_2$)}

In regime (III), which we call {\it strongly discrete}, the localization length $l$ of the DGB solutions is the smallest possible over the whole gap (i.e. is of the same order as $\xi$) and
$\max \frac{|u_{i}-u_{i+1}|}{\delta_{0}}\gg 1$. The LA-DGB is subject only to weak oscillatory instabilities while the HS-DGB continues to be subject to the real instability, which has been considerably strengthened. 

\begin{figure}[htbp]
\begin{center}
\includegraphics[width=12cm,height=8cm,angle=0,clip]{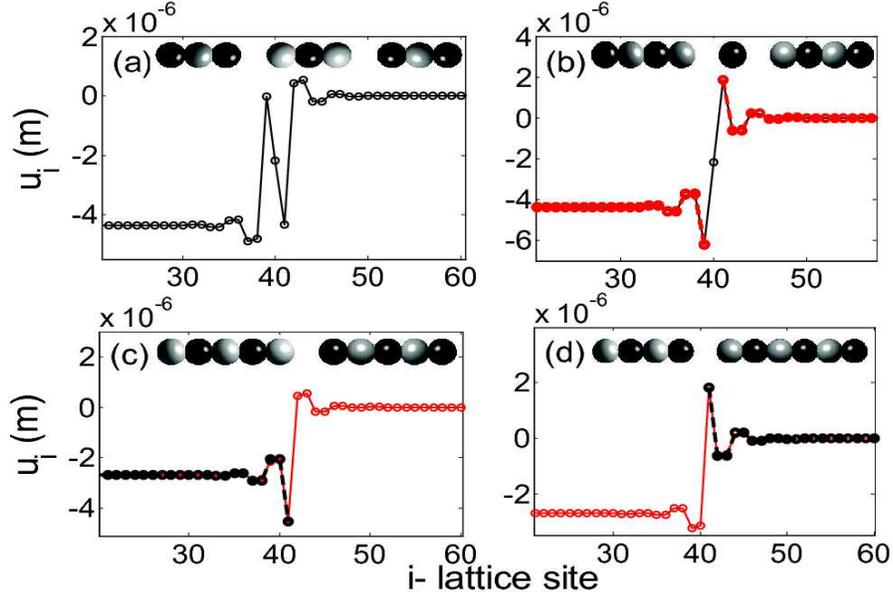}
\end{center}
\caption{(Color online) Top panels: Spatial profile of an HS-DGB with frequency $f_b=7210$ Hz at $t=0$ (a) and at $t=T/2$ (b). 
Bottom panels: As with the top panels, but for LA-DGB solutions. The dashed curves correspond to the spatial profile of the surface mode obtained using Eqs.~(\ref{e:left1BC},\ref{e:right1BC}). In each panel, we include a visualization of particle positions, and gap openings, for the corresponding time and DGB solution.}
\label{HA_LS_7000}
\end{figure}

The spatial profile of the DGB solutions
is now quite different from those in regimes (I) and (II) and hence from those of DGBs in a standard diatomic FPU-like system \cite{Maniadis2003}. In Fig.~\ref{HA_LS_7000}, we show examples of both families of DGBs at $f_b=7210$ Hz, the characteristic frequency of the linear surface mode (see Table ~\ref{table_th_vs_exp_linear}), at $t=0$ and $t=T/2$.

Comparing these solutions to their siblings in regimes (I) and (II) (for example, comparing Fig. 9a to Fig. 5a and Fig. 9c to Fig. 7a) reveals a remarkable change
in their spatial profiles. We observe, in addition to their more narrow profile, that for both families of DGB at this particular frequency, near the center of the DGB there exist adjacent pairs with small relative displacements (they move together). This qualitative shape is now reminiscent of the linear surface modes in Eqs.~(\ref{e:left1BC},\ref{e:right1BC}) instead of the spatial profiles characterizing the DGB in the other regimes. 

A possible explanation for the change of the spatial profiles of the DGBs is the following.  Near the center of the strongly discrete DGBs, we find that
\begin{equation}
	\frac{|u_{i-1}-u_{i}|}{\delta_{0}} \gg 1\,.
\label{approx1}
\end{equation}

From a physical perspective, this means that there is a large amount of time (in contrast to what we observe in the moderately discrete regime) during which
some beads near the center of the DGB lose contact with each other due to the tensionless Hertzian potential.  The system thus experiences effectively free boundaries conditions in the bulk as new "surfaces" are temporarily generated near the center of the DGB. However, as we have already mentioned for this type of system, a surface mode with a frequency in the gap (and now near this regime of DGB) exists only for free boundary conditions with light mass (aluminum) end particles. We observe, that for certain portions of the period of both HS and LA-DGB families, these conditions supporting a gap surface mode are satisfied. For this reason, in Fig.~\ref{HA_LS_7000}(b,c,d), we plot the spatial profile of the surface mode using Eqs.~(\ref{e:left1BC},\ref{e:right1BC}) and a corresponding visualization of our system which shows the location of gap openings and the newly created boundaries. For portions of the chain which have a light mass particle at the newly generated surface we overlay the displacement profile of the linear gap surface mode, with amplitude {\it B} of Eqs.~(\ref{e:left1BC},\ref{e:right1BC}) fitted to match the displacement of the DGB solution. 

At the particular frequency where the linear surface mode exists (see Fig. \ref{HA_LS_7000} , Eqs.~(\ref{e:left1BC},\ref{e:right1BC})
and associated discussion), we can observe the following phenomenology. As gap openings arise, 
there is a very good agreement between the surface mode displacement profile and the corresponding portion of DGB solution. 
On the other side of the chain (by necessity terminating in a heavy
particle), the waveform cannot form such a  surface gap mode. 
Importantly, the reader should be cautioned that this is a dynamical process during the oscillation period of the
``composite'' (of the above chain parts) breather where the gap openings arise and disappear during different fractions
of the breather period. Furthermore, it should be indicated that if the frequency deviates from the 
frequency of a linear surface gap mode, this phenomenology persists with the sole modification being
that instead of the linear surface gap mode, during gap openings, we observe a nonlinear variant thereof with a
progressively modified spatial profile.


Computation of the Floquet spectrum associated with linear stability shows that the HS-DGB modes in this regime
continue to be subject to the strong real instability. Additionally, the HS-DGB and the LA-DGB modes each possess $7$ quadruplets of eigenvalues that have left the unit circle. The maximum magnitude of these unstable Floquet multipliers is only about $1.02$, so the corresponding instabilities are very weak.  In order to address the question of how such instabilities manifest, we perform long-time simulations using as an initial condition the numerically exact LA-DGB with frequency $f_b=7000$Hz, which we perturb with white noise whose amplitude is $10\%$ of that of the LA-DGB.  The final result is the destruction of the DGB at $t \approx 0.075$ seconds soon following the generation of new internal frequencies and corresponding increase in the background noise. Thus, the corresponding LA-DGB has a finite lifetime of about $525 T$, where $T=1/f_b$ is the period of the breather.  This long-time evolution is reminiscent of that observed when DNLS single-site breathers are destroyed by standing-wave instabilities \cite{Johansson2000}.  

\subsection{Region (IV): Close to and Slightly Inside the Acoustic Band}

\begin{figure}[htbp]
\begin{center}
\includegraphics[width=8.5cm,height=7.5cm,angle=0,clip]{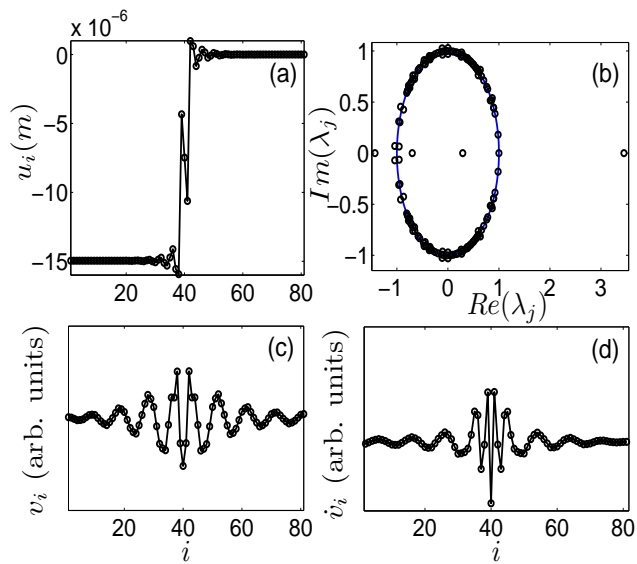}
\end{center}
\caption{(Color online)(a) Spatial profile of an HS-DGB with frequency $f_b=5500$ Hz. (b) Corresponding locations of Floquet multipliers $\lambda_{j}$ in the complex plane.  We show the unit circle to guide the eye. (c) Displacement and (d) velocity components of the Floquet eigenvectors associated with the second real instability (which, as described in the text, is a subharmonic instability). 
}
\label{HA_5500}
\end{figure} 

Finally, in regime (IV), which is close to and slightly inside
the acoustic band, both DGB solutions are delocalized (which implies that $l\gg \xi$) due to resonance with the upper acoustic mode while the amplitude $\max \frac{|u_{i}-u_{i+1}|}{\delta_{0}}\gg 1$. In this regime, 
both solutions are subject to strong oscillatory instabilities, and the HS-DGB solutions are still subject to strong real instabilities. 
However, there is also an interval---specifically from $f_b\approx 5940$ Hz to $f_b \approx f_1$---in which the HS-DGB family of breathers is subject to a second real instability.  This,
so-called, subharmonic instability is caused by the collision of a quadruplet of unstable Floquet multipliers at the $(-1,0)$ point on the unit circle. One pair of unstable Floquet multipliers returns to the unit circle but the second remains
on the real axis. We show the displacement and velocity components of the
associated unstable Floquet eigenvector at $f_b=5500$Hz in the bottom panels of Fig.~\ref{HA_5500}. As one can see, the displacement and velocity components of the Floquet eigenvectors are extended. Perturbing the solution along this subharmonic instability eigendirection and focusing only on short term dynamics
to avoid the manifestation of the stronger real instability, we observe a period doubling bifucation (oscillations of the central bead with twice the period of that of the oscillations of the adjacent beads).
  
\begin{figure}[htbp]
\begin{center}
\includegraphics[width=8.5cm,height=4cm,angle=0,clip]{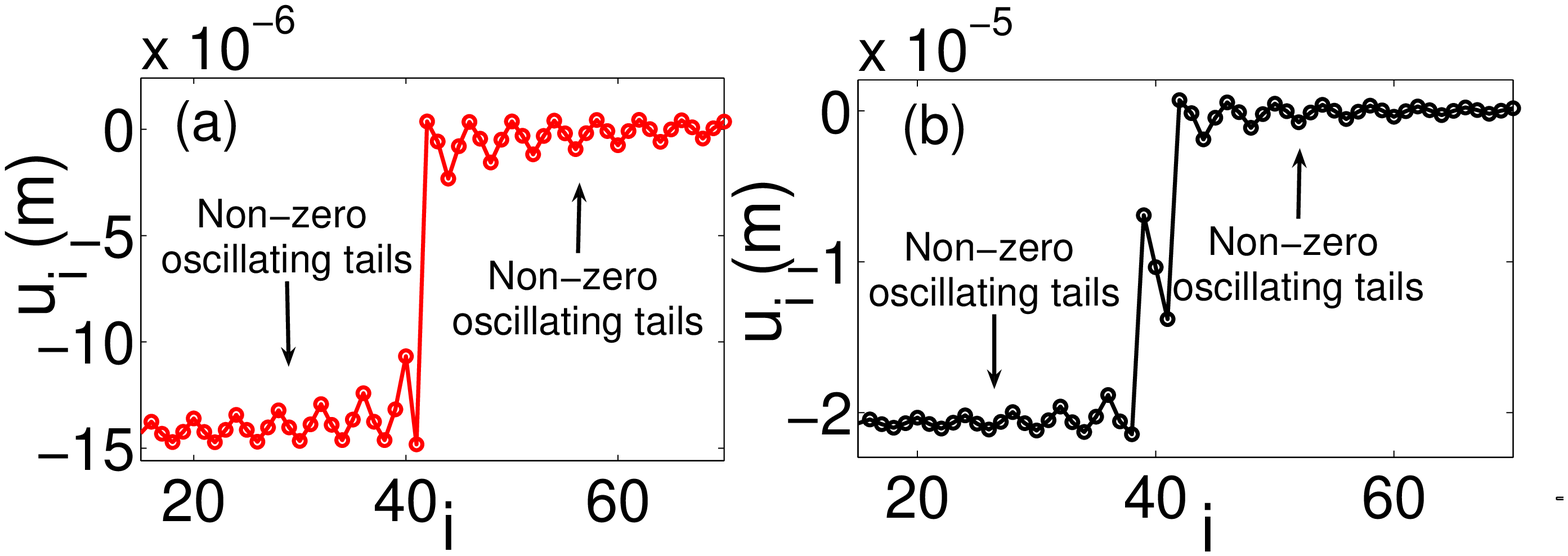} \\
\includegraphics[width=10cm,height=7cm,angle=0,clip]{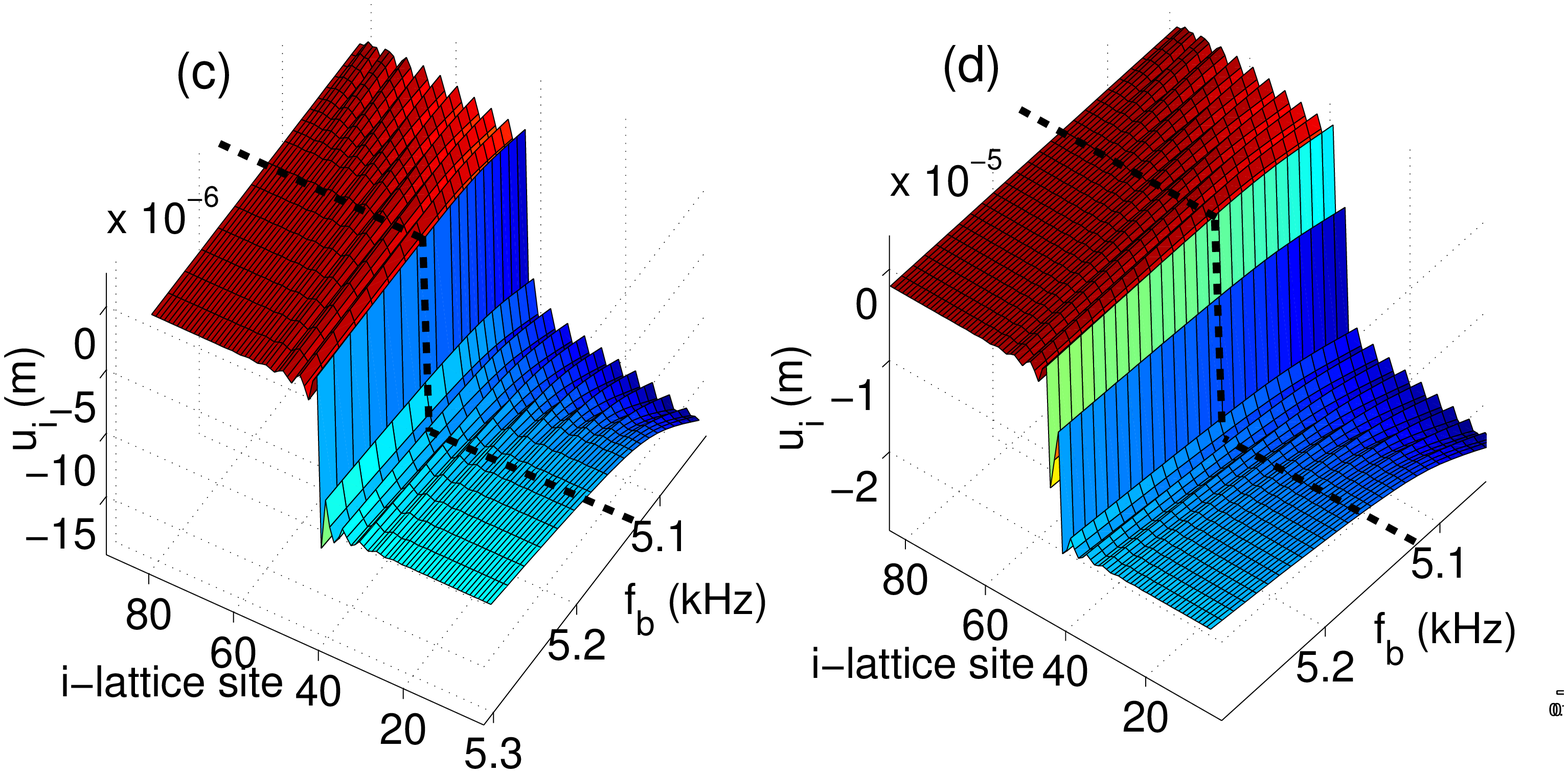}
\end{center}
\caption{(Color online) Spatial profile of a LA-DGB (a) and an HS-DGB (b) with frequency $f_b=5210$ Hz.
(c,d) Continuation of the DGBs into their discrete out gap siblings as the frequency crosses the
upper end of the acoustic band (denoted by dashed lines). The delocalization of the solution
profile as the upper acoustic band edge is crossed is evident for both the LA-DGB solutions (c)
and the HS-DGB solutions (d). 
}
\label{DOGB}
\end{figure}

Finally, we examine what happens at and slightly inside the acoustic band.  In contrast to the optical gap boundary, at which the DGB solutions delocalize and then vanish, we find in the acoustic boundary of the gap that the solutions delocalize, but persist with the addition of non-zero oscillating tails (see top panels of Fig.\ref{DOGB}).  These arise from resonance of the DGBs with the upper acoustic cutoff mode. The new bifurcated solutions are called \emph{discrete out-gap breathers} (DOGBs).  More about DOGBs and their possible bifurcations in a binary  DNLS model can be found in Ref.~\cite{Gorbach2004}.  In Fig.~\ref{DOGB}, we show the profiles of both families of DOGBs. In both cases, the DGBs transform into DOGB solutions with non-zero tails that have the form of the upper acoustic cutoff mode. The appearance of such modes, which are associated with resonances of the DGBs with the linear mode, can occur in general in finite-size systems in which the phonon spectrum is discrete.  They can be observed when the DGB frequency (or one of its harmonics) penetrates the phonon band. Other kinds of DGB solutions with different non-zero tails are generated when the second harmonic of the DGB penetrates the optical band from above. These solutions, which are called \emph{phonobreathers} \cite{phononDB}, have tails of the form of the optical upper cutoff mode and oscillate at a higher frequency (of about $f_3$).

\section{Conclusions}\label{sec5}

In this work, we have presented systematic computations of the intrinsically localized excitations that diatomic granular crystals can support in the gap of its spectrum between the acoustic and optical band of its associated linearization (linearizable under the presence of static compression). We have examined two families
of discrete gap breather (DGB) solutions.
One of them consists of heavy-symmetric DGBs (HS-DGB), and the other consists of light-asymmetric DGB (LA-DGB), where the symmetric/asymmetric characterization arises from the spatial profiles of their energy distributions.  We found that the HS-DGB branch of localized states is always unstable through the combination of an omnipresent real Floquet-multiplier instability and occasional oscillatory instabilities. We showed that the LA-DGB solutions have the potential to be stable as long as their frequency lies sufficiently close to the optical band edge. For lower frequencies, we observe within a small frequency interval that a real instability and also more broadly weak oscillatory instabilities render this solution weakly unstable, although in this case the solutions still might be observable for very long times.  We explored the progressive localization of the solutions upon decreasing the frequency within the gap, and we discussed the regimes of weak, moderate, and strong discreteness at length. We showed a unique spatial profile of DGB with strong discreteness, and their similarity to linear gap surface modes. Finally, in a specific frequency interval near the acoustic band edge of the linear gap, we also found a period-doubling bifurcation and described its associated instability. In the future it should be interesting to explore whether additional families of DGBs (including solutions that do not bifurcate from the linear limit) can exist in 1D or higher dimensional granular crystals.
%


\section{Acknowledgements} 

We thank Michael Weinstein and Vassilis Koukouloyannis for useful discussions. 
This work has been supported from ``A.S. Onasis" Foundation, RZG 003/2010-2011 (GT and PGK).
PGK gratefully acknowledges support from NSF-DMS-0349023 (CAREER) and NSF-DMS-0806762, as well
as from the Alexander von Humboldt Foundation. CD acknowledges support from the Army Research Office MURI 
(Dr. David Stepp) and from the National Science Foundation NSF-CMMI-0844540 (CAREER).


\end{document}